\documentclass[reprint,prl,amsmath,amssymb,aps]{revtex4-2}

\usepackage{graphicx}
\usepackage{dcolumn}
\usepackage{bm}
\usepackage{xcolor}
\usepackage[hidelinks]{hyperref}
\usepackage[none]{hyphenat} 

\graphicspath{{./Figures/}}

\begin{document}

\preprint{APS/123-QED}

\title{Thickness-Dependent Band Gap Modification in BaBiO$_{3}$}

\author{Rosa L. Bouwmeester}
\author{Alexander Brinkman}
\author{Kai Sotthewes}
 \affiliation{Faculty of Science and Technology and MESA+ Institute for Nanotechnology, University of Twente, 7500 AE Enschede, The Netherlands}
 
 \date{\today}
 
\begin{abstract}
The material BaBiO$_{3}$ is known for its insulating character. However, for thin films, in the ultra-thin limit, metallicity is expected because BaBiO$_{3}$ is suggested to return to its undistorted cubic phase where the oxygen octahedra breathing mode will be suppresse as reported recently. Here, we confirm the influence of the oxygen breathing mode on the size of the band gap. The electronic properties of a BaBiO$_{3}$ thickness series are studied using \textit{in-situ} scanning tunneling microscopy. We observe a wide-gap ($E_\textrm{G}$~$>$~1.2~V) to small-gap~($E_\textrm{G}$~$\approx$~0.07~eV) semiconductor transition as a function of a decreasing BaBiO$_{3}$ film thickness. However, even for an ultra-thin BaBiO$_{3}$ film, no metallic state is present. The dependence of the band gap size is found to be coinciding with the intensity of the Raman response of the breathing phonon mode as a function of thickness.
\end{abstract} 

\maketitle
The material system BaBiO$_{3}$ (BBO) received a lot of attention since it was first fabricated in 1963~\cite{Scholder1963} and is currently applied in various fields as in photoelectrochemical water splitting processes~\citep{Ge2018} and as an absorber in solar cells~\citep{Chouhan2018}. In a simple ionic picture, a Bi$^{4+}$ ion is expected to be present. Its half-filled 6\textit{s} shell would make BBO metallic~\cite{Pei1990, Baumert1995, Sleight1975, Cox1976}. Contrary, experimental results showed an insulating character~\cite{Cox1976, Tajima1985, Wertheim1982} with an optically observed bulk band gap of 2~eV ~\cite{Lobo1995, Tajima1987, Sato1989}, raising the question what mechanism is responsible for its insulating behavior. When more studies were performed, it was discovered that BBO became superconducting upon hole doping \cite{Pei1990, Baumert1995, Sleight1975, Sato1989, Mattheiss1988, Cava1988}. With optimum lead or potassium doping, critical temperatures of 13~K \cite{Sleight1975} and 30~K \cite{Cava1988}, respectively, were achieved. It is still an open question why BaPb$_{1-x}$Bi$_{x}$O$_{3}$ and Ba$_{1-x}$K$_{x}$BiO$_{3}$ become superconducting. A possible explanation might be related to the mechanism causing the unexpected insulating behavior in BBO. In this work, the electronic properties are studied to shine a new light on the mechanism responsible for the insulating character.

At first, a charge disproportionation, with alternating 3+ and 5+ valence states of the Bi atom, was given as the origin for the insulating behavior~\cite{Cox1976, Lobo1995,  Cox1979, Franchini2010}. Subsequently, others claimed the oxygen breathing mode~\cite{Tajima1985, Tajima1987, Mattheiss1982, Mattheiss1983} -- where the oxygen octahedra contract and expand -- to be responsible for the band gap formation in BBO. In a more recent view, a strong hybridization between the Bi~6\textit{s} and O~2\textit{p} states creates a bond disproportionation~\cite{Korotin2012, Foyevtsova2015, Khazraie2018}. Here, all bismuth ions have an oxidation state of 3+~\cite{Khazraie2018, Dalpain2018}, but different local environments. A hole pair sits on the contracted oxygen octahedra~\cite{Khazraie2018}, explaining the experimentally observed two different Bi-O bond lengths~\cite{Cox1976, Lobo1995, Cox1979}. In experimental work, the presence of an oxygen 2\textit{p} hole density in the ground state was observed for BBO single crystals~\cite{Balandeh2017} and a Bi core level analysis of BBO thin films revealed the presence of solely Bi$^{3+}$ ions~\cite{Plumb2016}, both agreeing with the theoretically proposed bond disproportionated picture.

\begin{figure}
\centering
\includegraphics[width=\linewidth]{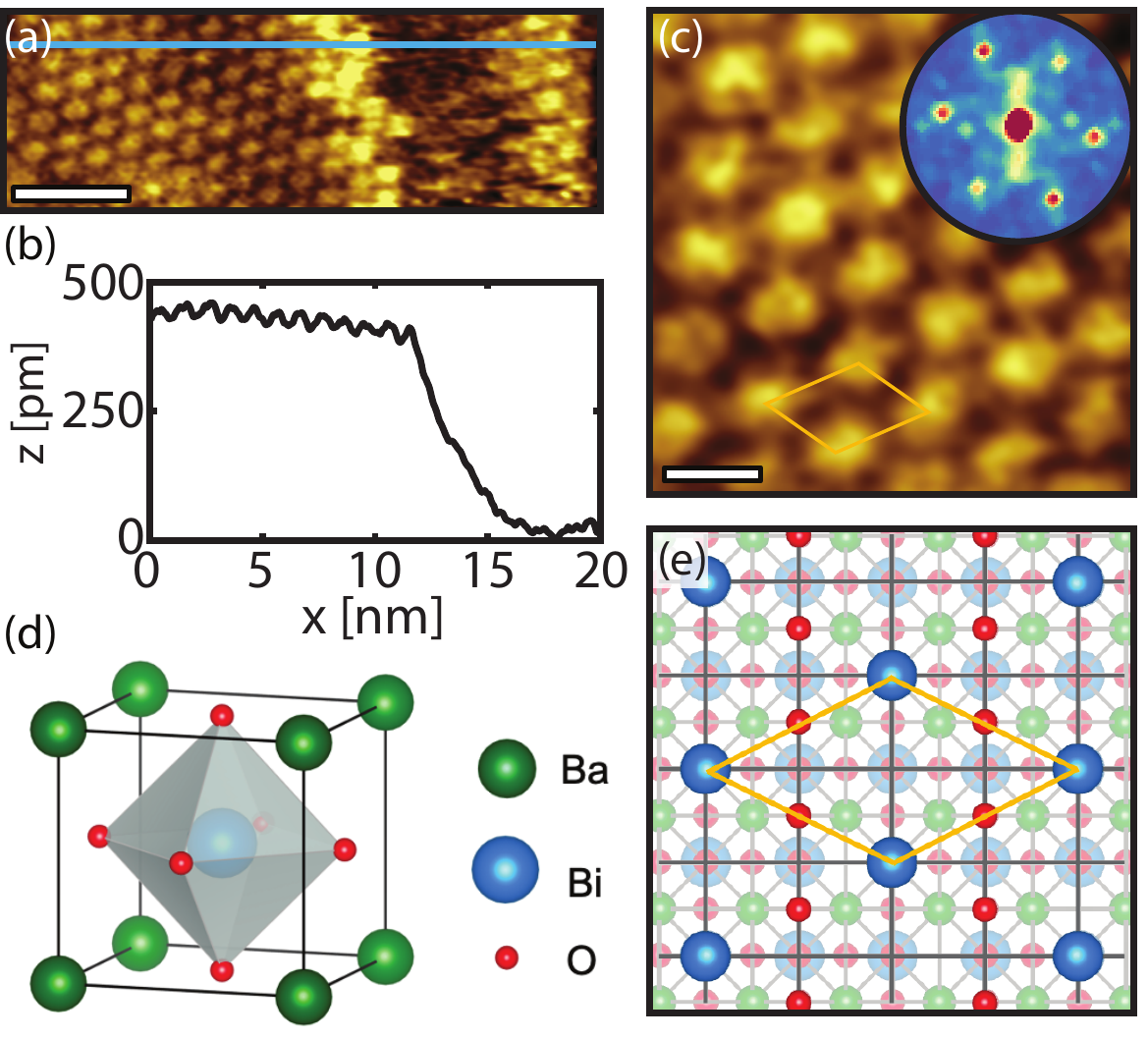}
\caption{(a) Topography image (20 $\times$ 7~nm, scale bar 4~nm) of the 10-unit-cell-thick BBO film on Nb:STO showing a stepped surface. (b) Cross-sectional height profile along the corresponding light blue line segment in (a). The step height is approximately 4.5~\AA. (c) Zoomed image (5~$\times$~5~nm, scale bar 1~nm) showing the c(4 $\times$ 2) surface reconstruction, the diamond-shaped orange lines indicate the unit cell. Inset: The corresponding FFT showing a threefold symmetry with a periodicity of 1~nm. (d) Schematic of the cubic perovskite BBO unit cell. The gray-shaded area represents an oxygen octaheder. (e) A top view of the c(4~$\times$~2) surface reconstruction. The diamond-shaped orange lines show the unit cell (corresponding to the orange diamond shape in (c)).}
\label{BBOTopo}
\end{figure}

In both the charge disproportionated and bond disproportionated picture, the oxygen octahedra breathing mode is present. The appearance of this breathing mode combined with octahedra tilting distortions cause a phase transition from a cubic to a monoclinic structure~\cite{Franchini2010}. When considering the thickness of the BBO films as a degree of freedom, it was found that for films thinner than 9 unit cells (u.c.) the structure transforms from a tetragonal to a cubic phase \cite{Kim2015}, hinting towards a suppression of the oxygen breathing mode. Theory shows a closing of the band gap when the oxygen octahedra breathing mode is excluded and a cubic BBO crystal structure is considered~\citep{Khazraie2018}. Nevertheless, no electrical properties were determined for any of these ultra-thin BBO films in order to confirm the presence or absence of a thickness-dependent insulator-to-metal transition.

\begin{figure}
\centering
  \includegraphics[width=\linewidth]{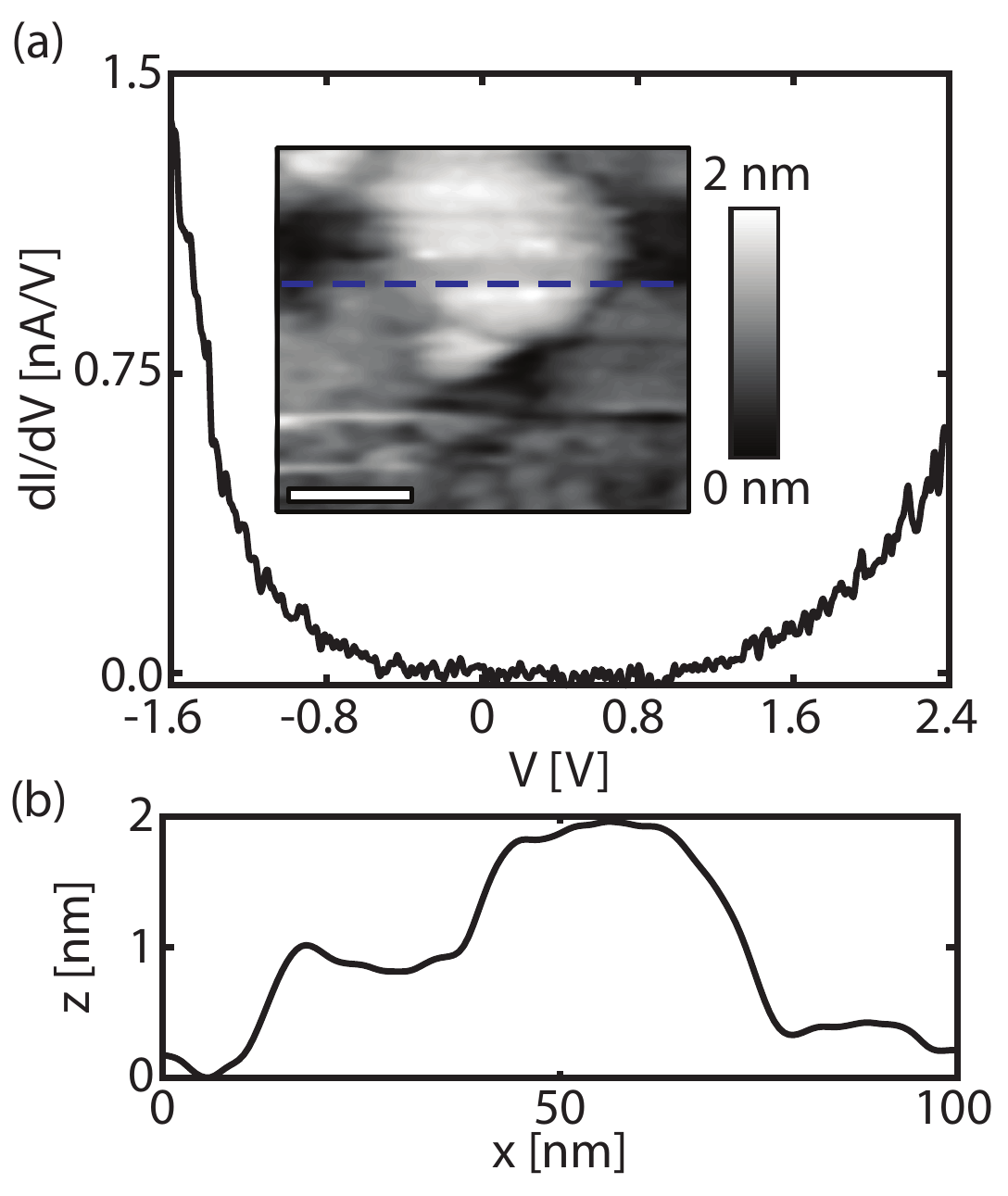}
\caption{Differential conductivity spectra ($\frac{\textrm{d}I}{\textrm{d}V}(V)$) on the 16-unit-cell-thick BBO film on Nb:STO(001). A band gap of approximately 1.2~$\pm$~0.3~eV is observed. Inset: Topography image (100~$\times$~100 nm, scale bar 30 nm). The tunneling parameters are 500~pA and -1.8~V. (b) Cross-sectional height profile of the corresponding blue dashed line in the inset of (a).}
\label{BBO16uc}
\end{figure}

Furthermore, with a Raman spectroscopy experiment on the same series of BBO films, a suppression of the breathing phonon intensity was observed as a function of thickness.  Repeating the experiment for BBO films deposited on a double buffer layer template of BaZrO$_{3}$ and BaCeO$_{3}$, a suppression of the breathing phonon intensity was not observed until 6~u.c.~\cite{Lee2018}. When BBO films were deposited directly on a Nb-doped SrTiO$_{3}$(001) substrate, the breathing phonon intensity was present till a thickness of 7~u.c.~\cite{Zapf2019}. However, it was concluded that the loss of response intensity was caused by a reconstruction layer at the interface (see refs. \cite{Zapf2018} and \cite{Bouwmeester2019}), rather than by suppression of the oxygen breathing mode. 

Here, we study the electronic properties of a BBO thickness series by \textit{in-situ} scanning tunneling spectroscopy (STS) experiments. We found that the size of the band gap ($E_\textrm{G}$) depends on the thickness of the BBO film and shrinks from $E_\textrm{G}$~$>$~1.2~V for a 16-unit-cell-thick film to $E_\textrm{G}$~$\approx$~0.07~eV for a film with a thickness of 3 unit cells. A c(4~$\times$~2) surface reconstruction confirms the presence of a perovskite structure.

\begin{figure*}
\centering
  \includegraphics[width=0.9\linewidth]{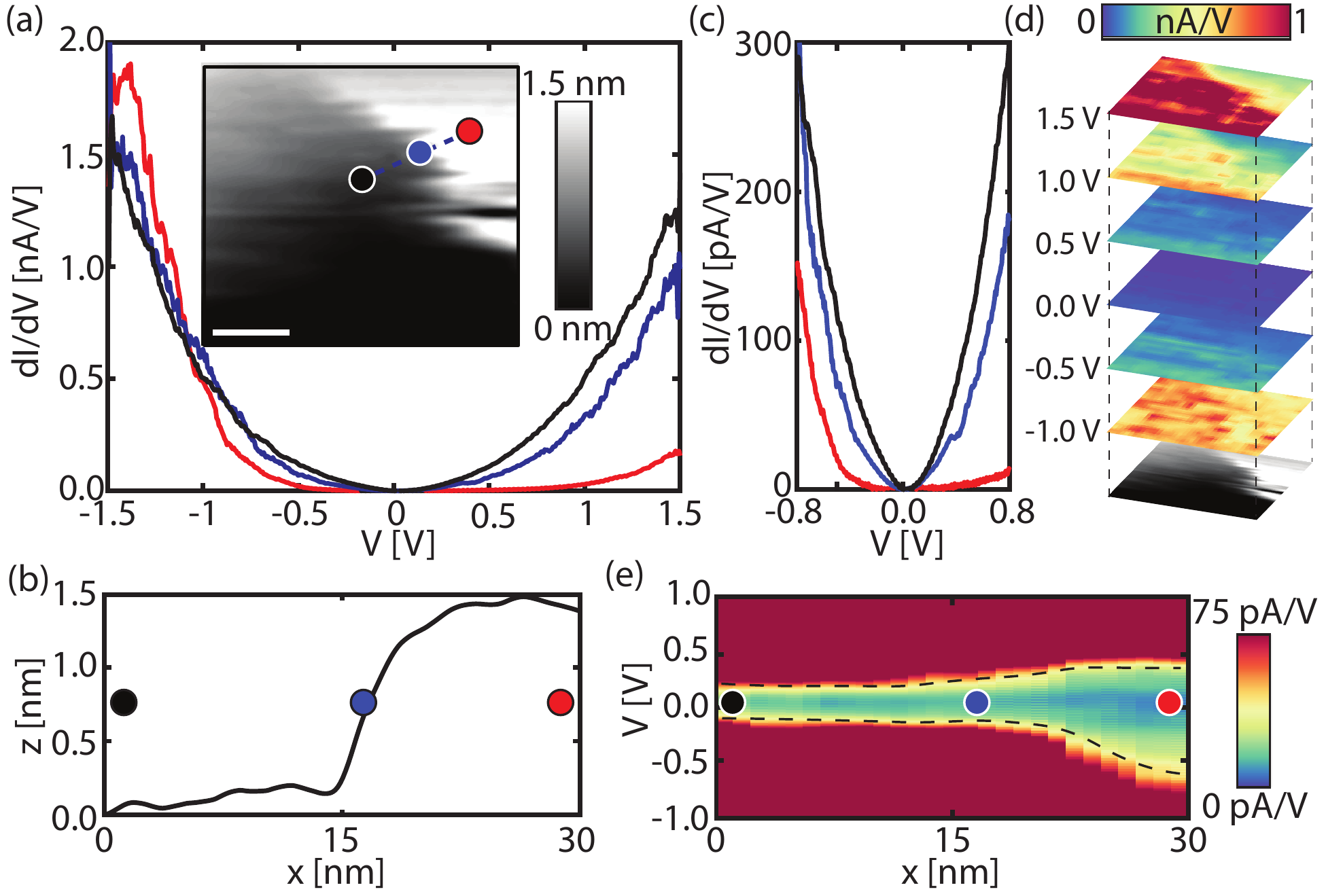}
\caption{(a) Differential conductivity spectra ($\frac{\textrm{d}I}{\textrm{d}V}(V)$) on the 10-unit-cell-thick BBO film on Nb:STO, the curve colors correspond to the positions (colored dots) marked in the inset. A wide band gap (red) to small band gap (black) transition is found depending on the thickness of the film. Inset: Topography image (100 $\times$ 100 nm, scale bar 25 nm) with spectrum locations marked by the colored dots. The tunneling parameters are 700 pA and -1.5 V. (b) Cross-sectional height profile, from corresponding blue dashed line in inset of (a), showing a height difference of 1.5~nm between the higher and lower region. (c) Zoom-in of the spectroscopy data of (a) revealing a decrease of the band gap when moving from the higher to the lower region. (d) d$I$/d$V$ maps at different bias voltage set points. The lateral position is aligned with the topography image at the bottom, which is the same as the inset in (a). An increased LDOS at the lower region of the sample is clearly visible for $V$~$>$~1~$V$. (e) $\frac{\textrm{d}I}{\textrm{d}V}(V)$ cross-section recorded across a transition region (colored dots correspond to the dots in the inset of (a)). A strong decrease of the band gap is observed when going from the higher (red dot) to lower (black dot) region. The black dashed line is a guide to the eye.}
\label{BBO10uc}
\end{figure*}

BBO films are fabricated with thicknesses ($d_{\textrm{BBO}}$) of 4, 10 and 16 u.c. on TiO$_2$-terminated, 0.5~wt\% Nb-doped SrTiO$_{3}$(001) substrates (Nb:STO) from CrysTec GmbH. The films were fabricated by pulsed laser deposition (PLD) using a stoichiometric house-made BaBiO$_3$ target (purity 99.99\%). The growth conditions of the BBO films were the same as reported in earlier work~\cite{Bouwmeester2019}. In the Supplementary Material, more details on sample preparation are provided together with the reflection high-energy electron diffraction (RHEED) patterns and intensity curves.

Subsequently, the BBO samples were transferred \textit{in-situ} to an Omicron Nanoprobe scanning tunneling microscope (STM), with a base pressure of 1~$\times$~10$^{-10}$ mbar using chemically etched tungsten tips. The measurements were acquired at room temperature. All voltages refer to the tip bias voltage with respect to the sample. The $\frac{\textrm{d}I}{\textrm{d}V}(V)$ curves were recorded using a lock-in amplifier ($f$~=~1763 Hz, $V_{\textrm{AC}}$ =~30~mV). In addition, $I(V)$ curves were recorded and used for calibration of the $\frac{\textrm{d}I}{\textrm{d}V}(V)$ curves. In order to compare the band gap obtained from the different samples, the curves are normalized with respect to each other. By plotting the corrected $I(V)$ spectra on a semi-logarithmic scale, the band gap is determined by taking the average voltage separation between the conduction and valence band current onsets at the lowest detectable current (detection limit approximately 500~fA)~\cite{Feenstra1987,Ebert2011,Herbert2013}, as explained in more detail in the Supplementary Material.

Fig.~\ref{BBOTopo}(a) shows the surface of the 10-unit-cell-thick BBO film on Nb:STO. The corresponding height profile across the surface is presented in Fig.~\ref{BBOTopo}(b) and reveals a step height of approximately 4.5~\AA. This is consistent with a single unit cell of BBO (\textit{a}~=~4.35~\AA) \cite{Bouwmeester2019}. Fig.~\ref{BBOTopo}(c) shows a zoomed image of the BBO surface. The observed pattern of atoms corresponds to a c(4~$\times$~2) surface reconstruction, the diamond-shaped orange lines indicate its unit cell. The same surface reconstruction is also found for the 4-unit-cell-thick BBO film, shown in the Supplementary Material. In Fig.~\ref{BBOTopo}(e) the surface reconstruction on the BBO surface is schematically depicted, the orange diamond-shaped lines indicate the unit cell and correspond to the orange lines in Fig.~\ref{BBOTopo}(c).

Erdman \textit{et.~al}~\cite{Erdman2003} concluded for the c(4~$\times$~2) reconstruction on a STO(001) surface that a stoichiometric TiO$_{2}$ overlayer was present, consisting of TiO$_{5}$ edge-sharing polyhedra. Such an overlayer is formed to stabilize the truncated, corner-sharing octahedra in the surface layer of a TiO$_{2}$-terminated STO(001) underneath. Therefore, we propose, in analogy with the studies on Nb:STO(001) and STO(001), that a BiO$_2$ overlayer is present on the surface of BBO, hosting BiO$_{5}$ edge-sharing polyhedra.

The Fast Fourier transform (FFT), shown in the inset of Fig.~\ref{BBOTopo}(c), has a threefold symmetry with a periodicity of 1~nm. Note that the underlying crystal structure is still a fourfold symmetric cubic perovskite structure. The lattice constant corresponding to a periodicty of 1~nm is 4.35~\AA~($\sqrt{5}~\times$~4.35~\AA~=~1~nm), which is in agreement with the BBO lattice constant. The surface reconstruction on the 4-unit-cell-thick BBO film also has a periodicity of 1~nm, as presented in the Supplementary Material. Previous studies on Nb:STO(001) and STO(001) surfaces revealed the same type of reconstruction, however, with a periodicity of 0.88~nm -- corresponding to a bulk lattice constant of 3.9~\AA~\cite{Erdman2003,Jiang1999,Castell2002} matching STO (\textit{a}~=~3.905~\AA)~\cite{Ohtomo2004}. 

The observation of BBO thin films with a relaxed lattice constant is in good agreement with our previous study~\cite{Bouwmeester2019}, where we show -- by means of a scanning transmission electron microscopy (STEM) --  that a lattice mismatch of 12\% between the STO substrate and BBO film is accommodated by the formation of an interfacial layer. This was confirmed by Jin \textit{et al.}~\cite{Jin2020}, where an interfacial layer with a fluorite structure and similar thickness is observed at the STO/BBO interface. The interfacial layer decouples the BBO film from the substrate. The bottommost part is still strained to the STO substrate, but the subsequent layer is already fully decoupled and relieves all strain in a dislocation every ninth unit cell~\cite{Bouwmeester2019}. Therefore, the substrate does not influence the BBO film.

\begin{figure}
\centering
  \includegraphics[width=\linewidth]{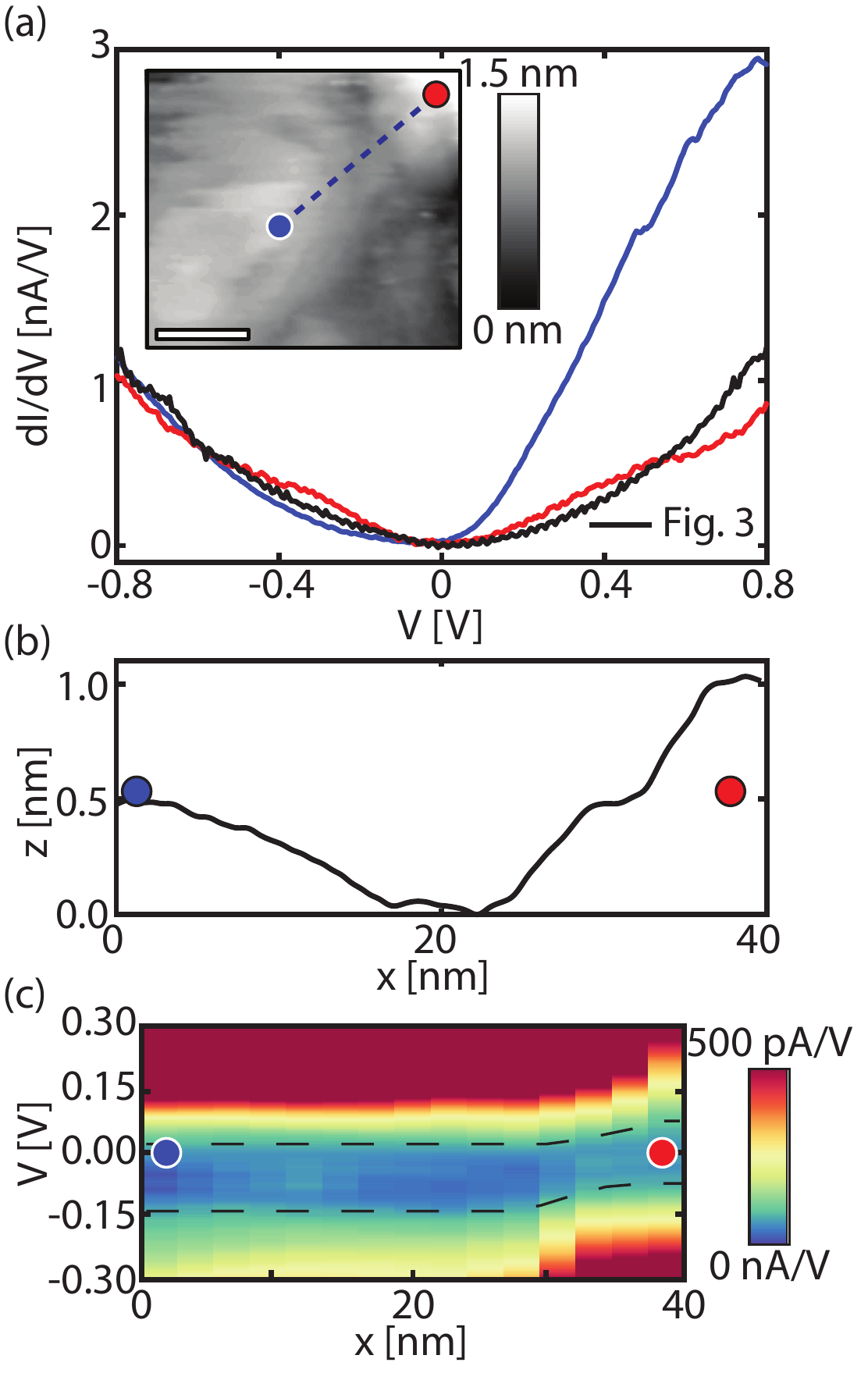}
\caption{(a) $\frac{\textrm{d}I}{\textrm{d}V}(V)$ spectra on the 4-unit-cell-thick BBO film on Nb:STO according to the positions (blue and red dot) marked in the inset. The black curve is taken from Fig. \ref{BBO10uc}(c) (the lower region) and rescaled based on the tunneling parameters. Inset: Topography image (80 $\times$ 80 nm, scale bar 20 nm) with the spectrum locations marked. The tunneling parameters are 400~pA and -1~V. (b) Cross-sectional height profile from corresponding blue dashed line in (a), a height variation of maximum 1~nm is observed. (c) $\frac{\textrm{d}I}{\textrm{d}V}(V)$ cross-section recorded across a transition region (blue dashed line in the inset of (a)).}
\label{BBO4uc}
\end{figure}

The BBO film is able to continue its growth without any effects of strain in a perovskite structure~\cite{Bouwmeester2019, Jin2020}. Anti-phase boundaries, where a step of half a unit cell is present, are observed within the BBO film in various STEM studies \cite{Zapf2018, Bouwmeester2019, Jin2020}. However, this boundary does not disturb the quality of the film. The suggested suppression of the oxygen breathing mode as a function of thickness has not yet been observed by STEM. The structural transition from tetragonal to cubic, as previously observed \cite{Kim2015}, is too small to be detected by STEM since it occurs on the picometer range.

The presence of the surface reconstruction, with an identical periodicity for the 4- and 10-unit-cell-thick BBO films, proves that the underlying structure in both cases is still the perovskite structure with the relaxed lattice constant of bulk BBO. The subsequently presented spatially-resolved STS measurements are, therefore, observing the effect of thickness and not of substrate-induced strain nor defects.

STS was first used to determine the electronic properties of the 16-unit-cell-thick BBO film. In Fig.~\ref{BBO16uc}(a), the differential conductance curves ($\frac{\textrm{d}I}{\textrm{d}V}(V)$) represent the local density of states (LDOS). A semiconducting characteristic with a band gap ($E_\textrm{G}$) of 1.2~$\pm$~0.3~eV is observed. From the cross-section height profile (see Fig.~\ref{BBO16uc}(b)) variations of 2~nm are observed, which have no clear spatial-dependent influence on the electronic profile across the surface. The band gap is asymmetric with the valence band located closer to the Fermi level ($E_\textrm{F}$).

In the case of the 10-unit-cell-thick BBO film, the LDOS varies spatially. Fig~\ref{BBO10uc}(a) shows the LDOS on various locations on the sample, the colored dots on the topography image in the inset correspond to colors of the differential conductance spectra. On the thicker part of the sample, represented by the red curve in Fig.~\ref{BBO10uc}(a), semiconducting behavior is observed -- similar to the profile on the 16-unit-cell-thick BBO film in Fig.~\ref{BBO16uc}(a). However, the size of the band gap is reduced to 0.7~$\pm$~0.2~eV.

From the cross-section height profile, see Fig.~\ref{BBO10uc}(b), it is clear that the lower region (black dot) is located 1.4~nm ($\approx$ 3~u.c.) lower with respect to the higher region (red dot). Note that the exact thickness of the BBO film is unknown and therefore only relative thicknesses are given. The $\frac{\textrm{d}I}{\textrm{d}V}(V)$ spectra taken at the lower region of the sample, black curve in Fig.~\ref{BBO10uc}(a), reveal a significant reduction of $E_\textrm{G}$.~A closer look at the Fermi energy, depicted in Fig.~\ref{BBO10uc}(c), shows this even more clearly. The band gap reduces to approximately 0.10~$\pm$~0.03~eV. 

The simultaneously obtained differential conductance (d$I$/d$V$) maps, presented in Fig.~\ref{BBO10uc}(d), depict the LDOS for the thickness regions at different bias voltages. A clear correlation is observed between the topography, bottommost image, and the LDOS. At non-zero bias voltages, the LDOS is much higher at the lower region of the sample compared to the higher region. At zero bias voltage all the contrast is lost and the regions can no longer be distinguished, excluding the presence of a metallic state.

In between the two height regions, a transition region is present, indicated by the blue dot in the inset of Fig.~\ref{BBO10uc}(a) and by the blue curve in (a) and (c). The size of the band gap depends heavily on the exact location on the sample, this is better visualized in Fig.~\ref{BBO10uc}(e). The measured $\frac{\textrm{d}I}{\textrm{d}V}(V)$ curves are plotted as a function of the distance, the colored dots correspond to the dots in the inset of Fig.~\ref{BBO10uc}(a). A continuous but steep transition in the band gap size is found between the two regions.

To further confirm the observed band gap reduction, the 4-unit-cell-thick BBO film is studied. Fig.~\ref{BBO4uc}(a) shows the location-dependent differential conductance. The LDOS measured on the higher region of the 4~u.c. BBO film (red curve in Fig.~\ref{BBO4uc}(a)) has the same characteristics as the spectrum measured on the lower region of the 10~u.c. BBO film (black curve in Fig.~\ref{BBO10uc}(a) and (c)), including a similar band gap of 0.10~$\pm$~0.05~eV. For better visualization, the scaled differential conductance spectrum of Fig.~\ref{BBO10uc}(c) (black curve) is also plotted in Fig.~\ref{BBO4uc}(a). 

On the region located 1~u.c.~lower, see the height profile in Fig.~\ref{BBO4uc}(b), the LDOS is altered again (blue curve in Fig.~\ref{BBO4uc}(a)). In addition to a small decrease in the size of the band gap (i.e. 0.07~$\pm$~0.04 eV), the band gap is no longer positioned symmetrically relative to the Fermi energy. A small shift towards negative voltages is observed (i.e. the conductance band is located closer to $E_\textrm{F}$), which is most likely caused by band bending. The contact potential difference between Nb:STO and BBO results in an accumulation layer of electrons in the BBO film, bending the conduction band towards the Fermi energy. Fig. \ref{BBO4uc}(c) shows the $\frac{\textrm{d}I}{\textrm{d}V}(V)$ as a function of the distance. The shift of the band gap towards more negative voltages at the lower region (blue dot) is clearly observed. 

From the BBO thin film thickness series, we observed the following: 1) a reduction of the band gap as a function of decreasing thickness, 2) the absence of a metallic state and 3) a monotonous but steep transition in the band gap size between the different thickness regions. Although an insulator-to-metal transition was predicted~\cite{Kim2015}, a clear wide-gap ($E_\textrm{G}$~$>$~1.2~eV) to small-gap ($E_\textrm{G}$~$\approx$~0.07~eV) semiconductor transition is observed as a function of $d_{\textrm{BBO}}$.

\begin{figure}
\centering
  \includegraphics[width=\linewidth]{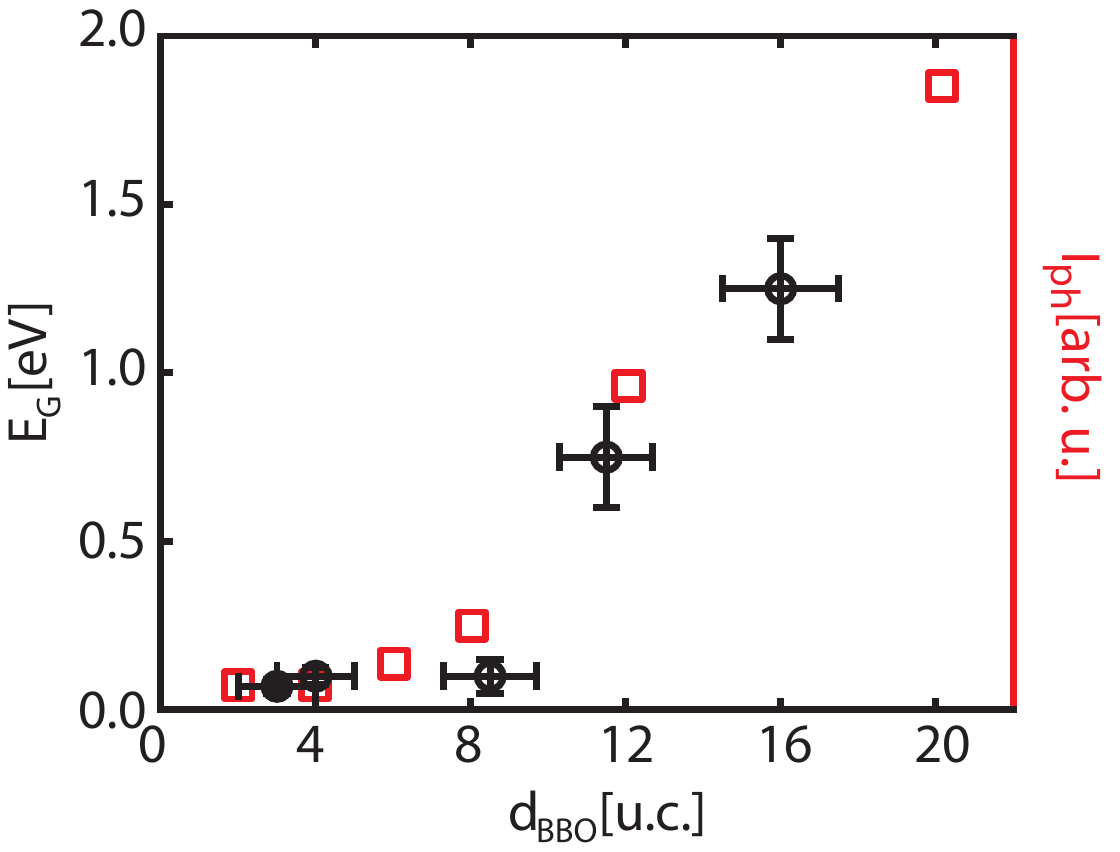}
\caption{The band gap (left axis, $\circ$) and the relative intensity of the Raman response of the breathing phonon mode ($I_\textrm{ph}$) (right axis, red squares, taken from ref.~\cite{Zapf2019}) as a function of the BBO film thickness ($d_{\textrm{BBO}}$). A similar trend is found for $E_\textrm{G}$ and $I_\textrm{ph}$. The x-axis error bar is the root-mean-square (RMS) roughness extracted from atomic force microscopy images (see Supplementary Material).}
\label{Fig:Bandgap}
\end{figure}

In Fig.~\ref{Fig:Bandgap}, the determined band gap sizes are plotted as a function of thickness. The uncertainty in thickness for the BBO films is determined from atomic force microscopy (AFM) images, presented in the Supplementary Material. The transition from a wide-gap to a small-gap semiconductor is continuous and gradual. Recent density functional theory (DFT) calculations in combination with a tight-binding (TB) model~\citep{Khazraie2018} revealed the influence of the oxygen breathing mode on the band structure of BBO. If the oxygen breathing mode is absent, a metallic band structure is predicted, while a band gap forms when the oxygen breathing mode is present. Implying that a semiconductor-metal transition occurs when the oxygen breathing mode is suppressed, as we observe as a function of thickness. No metallicity is observed, even for the 4-unit-cell-thick BBO film, implying that the oxygen breathing mode is not fully suppressed.

On the right axis of Fig.~\ref{Fig:Bandgap}, the intensity of the Raman response of the breathing phonon mode ($I_\textrm{ph}$) is plotted (data from ref.~\cite{Zapf2019}). A coinciding dependence between $E_\textrm{G}$ and $I_\textrm{ph}$ as a function of the BBO film thickness is observed. Synchrotron XRD and Raman spectroscopy experiments suggest that the strength of the oxygen breathing mode is decreasing with thickness \citep{Kim2015, Lee2018, Zapf2019}. Therefore, we conclude that the closing of the band gap originates from the suppression of the oxygen breathing mode as a function of thickness.

For increasing thickness, around a thickness of 8~u.c., both the band gap size and the intensity of the breathing phonon intensity start to increase. In both cases, the BBO films remain insulating up to the ultra-thin limit of $d_{\textrm{BBO}}=$~3~u.c. In addition, no step-like suppression in the breathing phonon intensity or the band gap is observed, in good agreement with previous studies \cite{Kim2015,Lee2018,Zapf2019}. The observed c(4~$\times$~2) surface reconstruction on the BBO films with thicknesses of 4 and 10 u.c., confirms the underlying structure is the perovskite structure and not an interfacial layer \cite{Bouwmeester2019, Jin2020} and the identical periodicity proves that no substrate-induced strain effect is present.

The influence of the breathing phonon mode on the band gap also explains the discrepancy between the observed band gap of the 16-unit-cell-thick BBO film (shown in Fig.~\ref{BBO16uc}) and the optical determined band gap on BBO (1.2~eV versus 2.0~eV) in previous studies (for single crystals \cite{Lobo1995, Tajima1987} and for films with $d_{\textrm{BBO}}$~$>$~300~nm~\cite{Sato1989}). This discrepancy is a consequence of the thickness-dependent band gap variation. In refs.~\cite{Kim2015,Zapf2019}, it is clearly shown that the BBO breathing phonon intensity was already affected for BBO thicknesses of 30~u.c. Therefore, the band gap on the 16-unit-cell-thick BBO film is already reduced and not reaching the optically measured 2~eV band gap. 

In conclusion, by combining STM and STS, we observed a thickness-dependent wide-gap to small-gap semiconductor transition in BBO thin films.~For $d_{\textrm{BBO}}\approx$~16 u.c., a band gap of 1.2~$\pm$~0.3~eV is observed. With a reduction of the film thickness, the band gap shrinks to approximately 0.07~$\pm$~0.04~eV for a 3-unit-cell-thick BBO film. No metallic state was detected in the ultra-thin limit. The transition is continuous and gradual and shows a coinciding dependence with the intensity of the Raman response of the breathing phonon mode as a function of thickness. A c(4~$\times$~2) surface reconstruction is observed on the 4- and 10-unit-cell-thick BBO films, confirming the perovskite structure with the correct lattice constant underneath, excluding the influence of substrate-induced strain. The presented results show that the suppression of the oxygen breathing mode as a function of thickness is responsible for the modification of the band gap size.
\\ \\

The authors thank Dominic Post and Martin Siekman for technical support and Prof. Dr. Ir. Harold J.W. Zandvliet for fruitful discussions.

\newpage
\renewcommand*{\thefigure}{S\arabic{figure}}
\setcounter{figure}{0}
\section{}

\maketitle

\newpage
\begin{widetext}
\section{\large{Supplementary Material: \\ \vspace{0.5cm} Thickness-Dependent Band Gap Modification in BaBiO$_{3}$}}
\end{widetext}

\section{Sample fabrication and characterization}

\begin{figure}
\includegraphics[clip,keepaspectratio,width=0.35\textwidth]{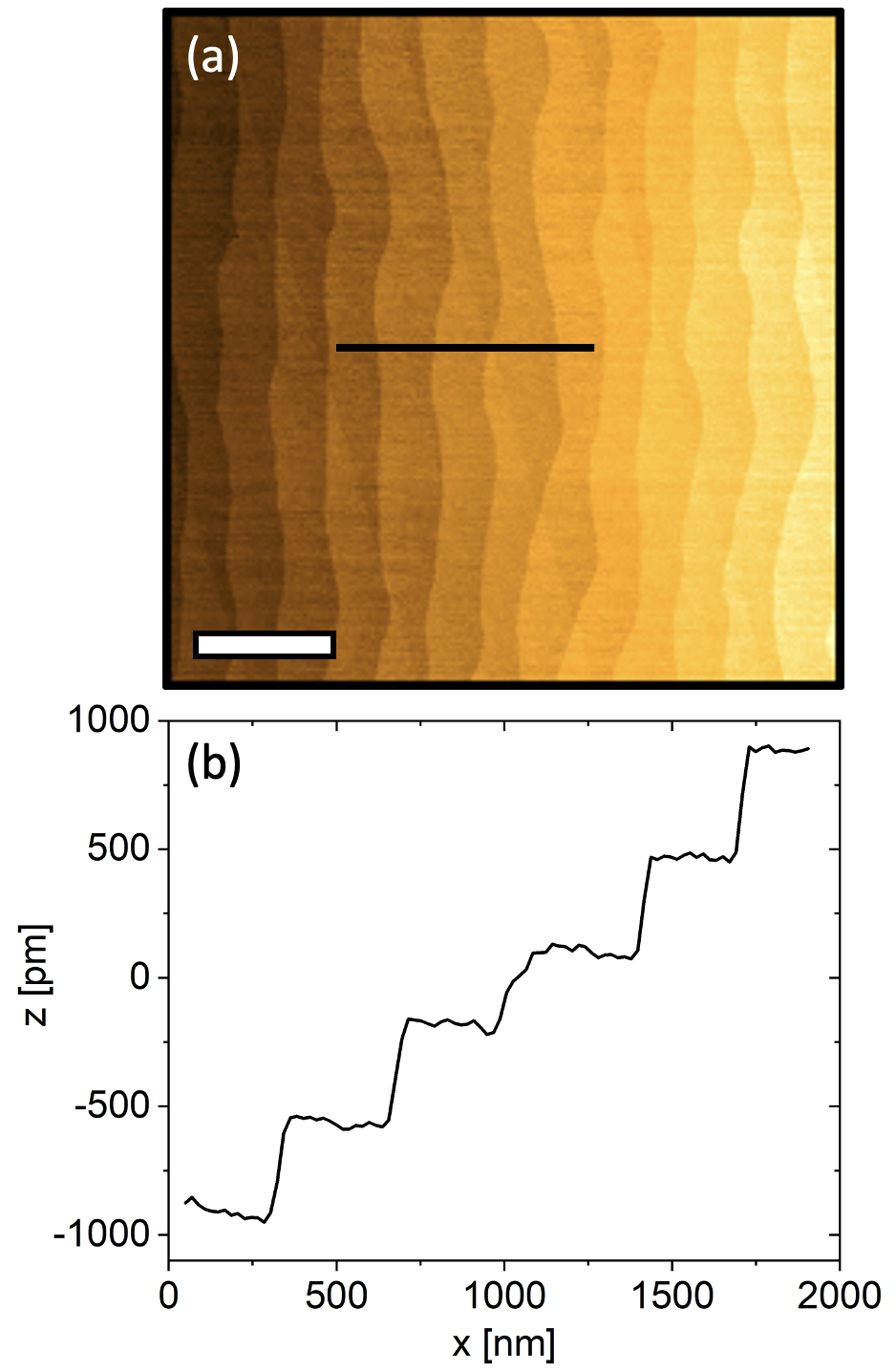}
\caption{(a) AFM image of a 4~$\times$~4~$\mu$m area on a Nb-doped STO(001) substrate. The root-mean-square (RMS) roughness is 168~pm. The scale bar is 1~$\mu$m. (b) The height profile corresponding to the black line in (a), confirming the TiO$_{2}$ single terminated surface.}
\label{Substrate}
\end{figure}

As a substrate Nb-doped SrTiO$_{3}$(001) (Nb:STO) (from CrysTec GmbH) was used, with a doping level of 0.5~wt\%. To obtain a TiO$_{2}$ single-terminated surface, a wet etching step of 30~seconds in a buffered hydrogen fluoride solution was performed \cite{Koster1998}. Subsequently, the substrates were annealed for 1.5~hours in a furnace with a continuous oxygen flow at 930~$^{\circ}$C. Afterwards, the surface quality was checked with an atomic force microscope (AFM). A 4~$\times$~4~$\mu$m image of a Nb:STO is presented in Fig.~\ref{Substrate}(a), nicely arranged terraces are observed. The height profile, see Fig.~\ref{Substrate}(b) and corresponding to the black line in (a), shows straight terraces with a step height of approximately 0.4~nm -- in good agreement with the STO bulk lattice constant of 3.905~\AA~\cite{Ohtomo2004}.

The BaBiO$_{3}$ thin films were fabricated using pulsed laser deposition (PLD). Before each deposition, the target was sanded and a pre-ablation was performed with 600~pulses fired at 5~Hz in the same oxygen background pressure as used during the actual deposition. The same growth conditions were used as in \cite{Bouwmeester2019}: a KrF laser at a fluence of 1.9 J/cm$^{2}$ with a repetition rate of 1~Hz, substrate temperature of 500~$^{\circ}$C, oxygen background pressure of 0.0100~mbar and a substrate-target distance of 50~mm.

\begin{figure}
\includegraphics[clip,keepaspectratio,width=0.5\textwidth]{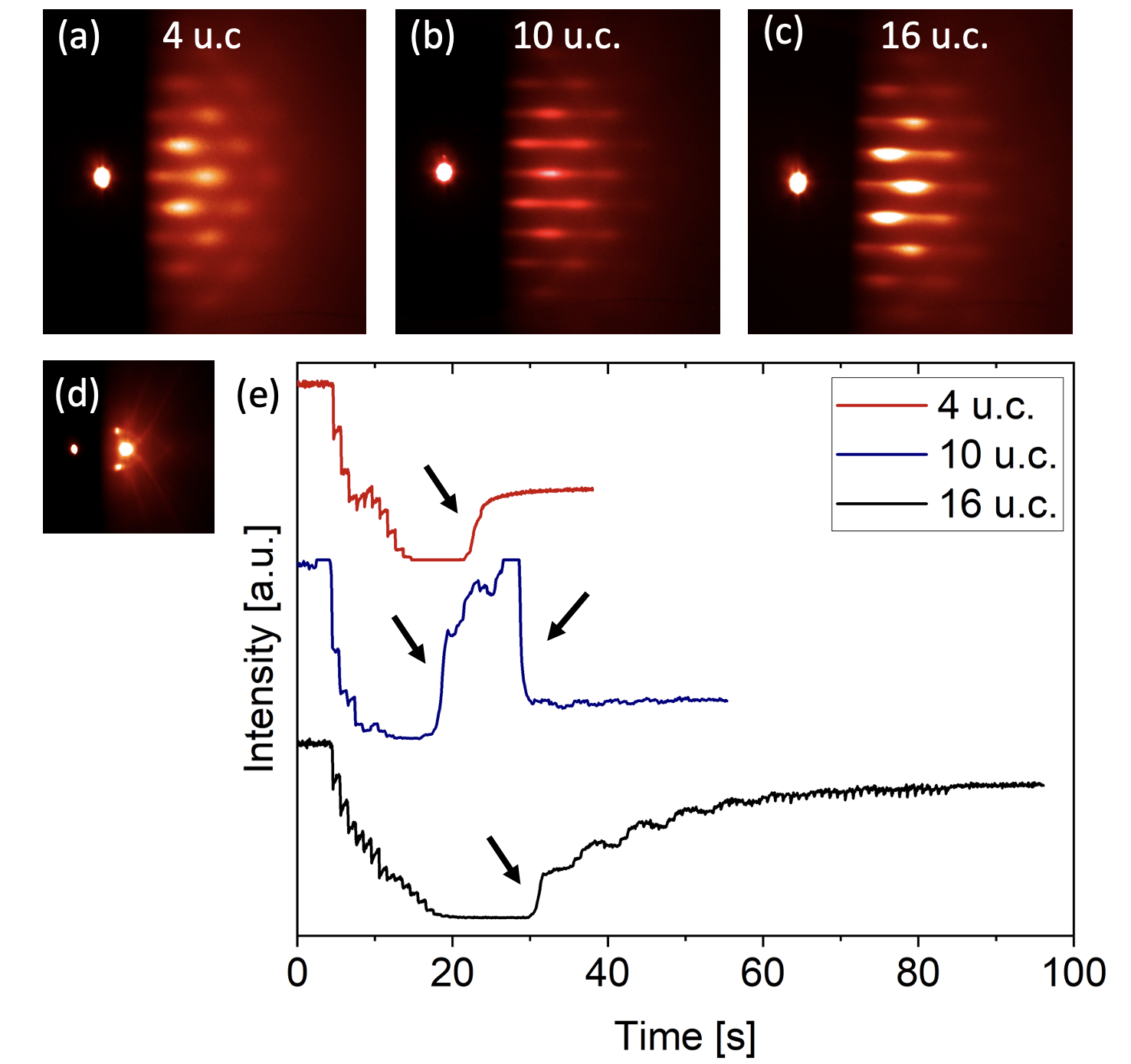}
\caption{(a-c) RHEED images of the diffraction patterns of the 4-, 10- and 16-unit-cell-thick BBO films, respectively. The images are taken after the samples were cooled down and in high vacuum conditions. (d) RHEED pattern of the Nb-doped STO(001) substrate used for the 4~u.c. BBO film, taken at room temperature in high vacuum conditions. (e)~The intensity of the main diffraction spot of the RHEED pattern is monitored during growth. The red, blue and black curves correspond to the 4-, 10- and 16-unit-cell-thick BBO films, respectively. The small black arrows indicate where the RHEED intensity was manually adjusted.}
\label{RHEED}
\end{figure}

During the deposition, the growth of the BBO films was monitored by reflection high-energy electron diffraction (RHEED). In Fig.~\ref{RHEED}(a-c) the RHEED patterns of the BBO films with thicknesses 4, 10 and 16~unit cells (u.c.), respectively, are presented. The images are taken after cool down in high vacuum conditions (average pressure of 3 x 10$^{-7}$~mbar). In Fig.~\ref{RHEED}(d) the diffraction pattern of the Nb:STO substrate, used for the 4-unit-cell-thick BBO film, is shown. In Fig.~\ref{RHEED}(e), the intensity of the main diffraction spot, that was monitored during growth, is presented as function of time. The red, blue and black curve correspond with the 4-, 10- and 16-unit-cell-thick BBO films, respectively. The small black arrows indicate where the intensity is manually de- or increased.

\begin{figure}
\includegraphics[clip,keepaspectratio,width=0.45\textwidth]{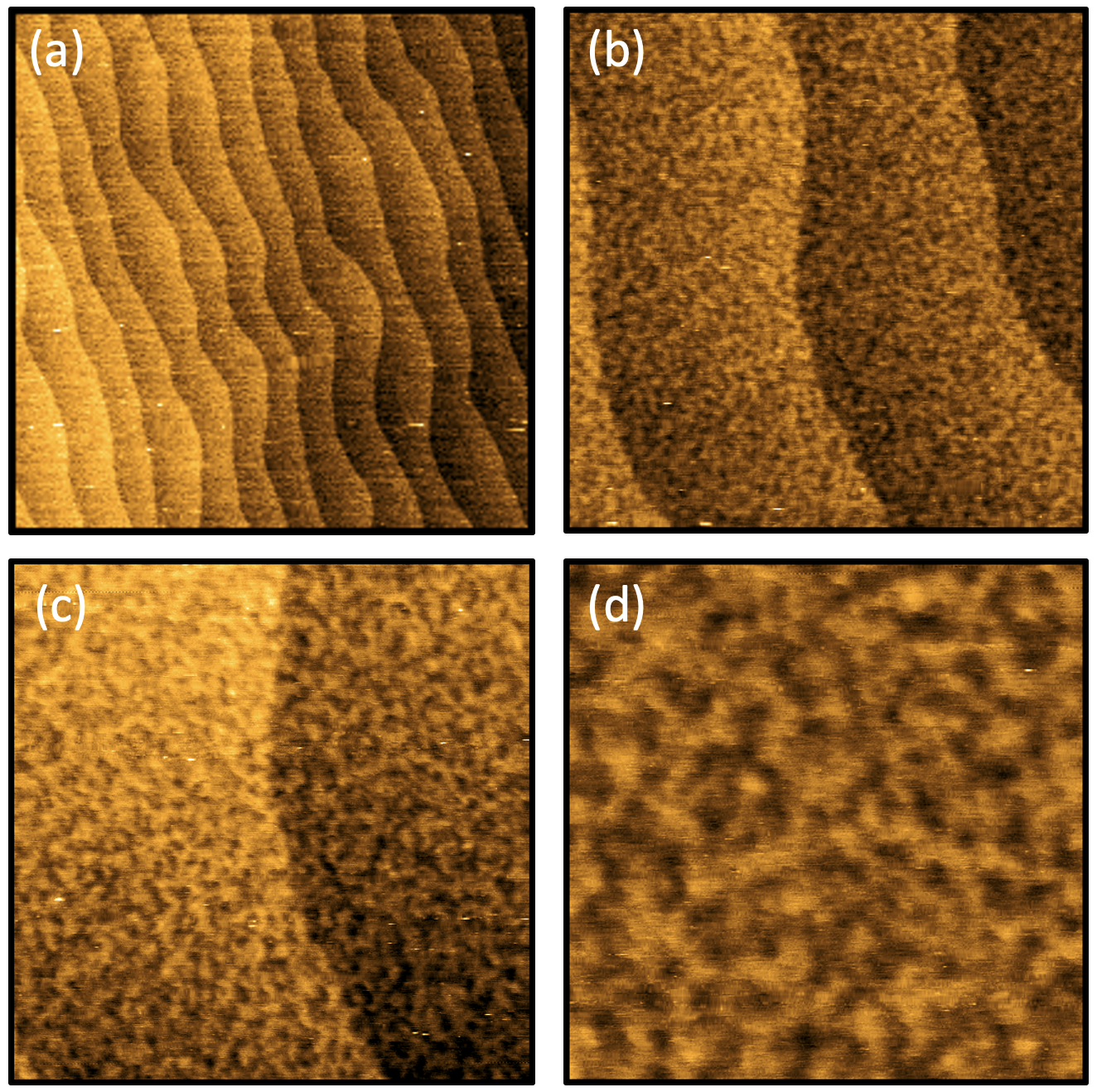}
\caption{AFM images of a 4-unit-cell-thick BBO films. The area is (a) 4~$\times$~4~$\mu$m, (b) 750~$\times$~750~nm, (c) 500~$\times$~500~nm and (d) 200~$\times$~200~nm. The RMS value is 150~$\pm$~20~pm.}
\label{AFM_4uc}
\end{figure}

The three BBO films were \textit{in-situ} transferred to a Nanoprobe scanning tunneling microscope (STM) for spectroscopy experiments. A 4-unit-cell-thick BBO film was removed from the vacuum and directly studied with an AFM, the result is presented in Fig.~\ref{AFM_4uc}. The substrate terraces are still visible, but when scanning a smaller area (the area decreases going from (a) to (d)) some roughness is observed. The root-mean-square (RMS) roughness is 150~$\pm$~20~pm. 

\begin{figure}
\includegraphics[clip,keepaspectratio,width=0.45\textwidth]{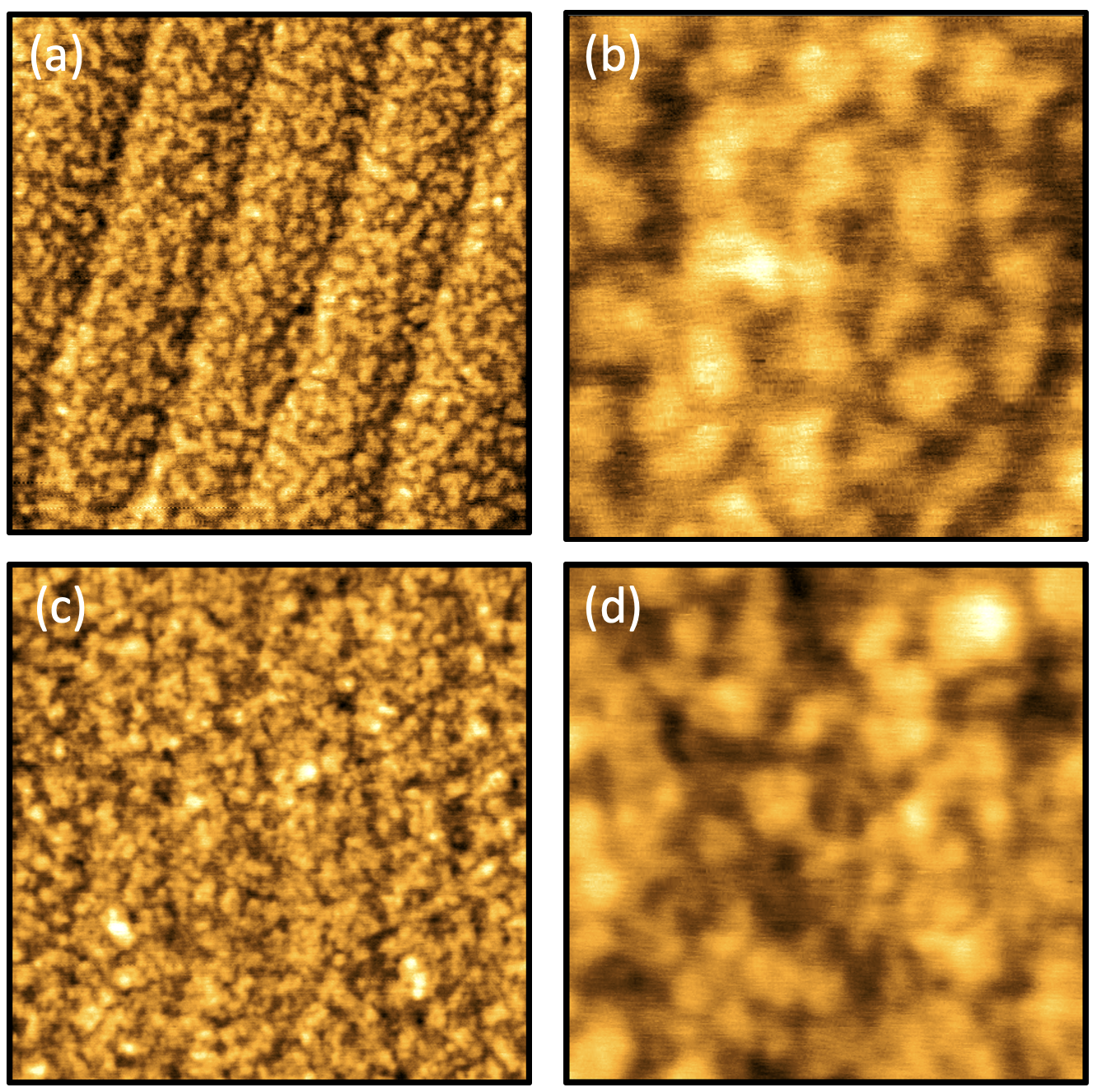}
\caption{AFM images of (a, b) a 10~u.c. and (c, d) a 16~u.c. BBO film. The area is (a, c) 750~$\times$~750~nm and (b,~d)~200~$\times$~200~nm. The RMS values for the 10- and 16-unit-cell-thick BBO films are 200~$\pm$~20~pm and 350~$\pm$~20~pm, respectively.}
\label{AFM_10uc}
\end{figure}

The same is performed for a 10- and a 16-unit-cell-thick BBO film (see Fig.~\ref{AFM_10uc}), the AFM images are presented in (a, b) and (c, d), respectively. The average RMS roughnesses are 200~$\pm$~20~pm and 350~$\pm$~20~pm for the 10 and 16~u.c. BBO films, respectively. The RMS values are used as a measure for the x-axis error bars in Fig.~5 of the main text.

\section{Scanning tunneling microscopy} 
The results of the spectroscopy experiments with the 4-, 10- and 16-unit-cell-thick BBO films are presented in the main text. Below, some additional results are presented.

\subsection{Surface reconstruction of a 4-unit-cell-thick film}
\begin{figure}
\includegraphics[clip,keepaspectratio,width=0.45\textwidth]{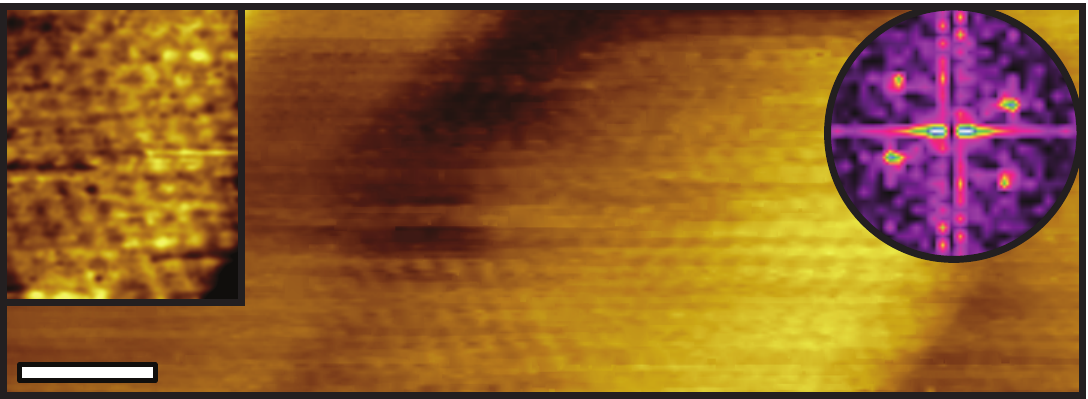}
\caption{Topography image (40~$\times$~13~nm, scale bar 5~nm) of the 4-unit-cell-thick BBO film on Nb:STO, showing the c(4~$\times$~2) surface reconstruction. The image is slightly distorted due to to the presence of a double tip. Inset left: Zoomed image (7.5~$\times$~9 nm) of the c(4~$\times$~2) reconstruction. Inset right: The corresponding Fast Fourier transform (FFT) showing the threefold symmetry with a periodicity of approximately 1~nm. The tunneling parameters are 700~pA and -1.5~V.}
\label{4ucReconstruction}
\end{figure}

Fig.~\ref{4ucReconstruction} shows the surface of the 4-unit-cell-thick BBO film on Nb:STO. The image is slightly distorted due the presence of a double tip. The atoms are arranged in the same symmetry and periodicity as for the 10-unit-cell-thick BBO films (Fig.~1 of the main text), corresponding to a c(4~$\times$~2) surface reconstruction.

\subsection{LDOS 4 u.c. BBO film}
\begin{figure}
\includegraphics[clip,keepaspectratio,width=0.25\textwidth]{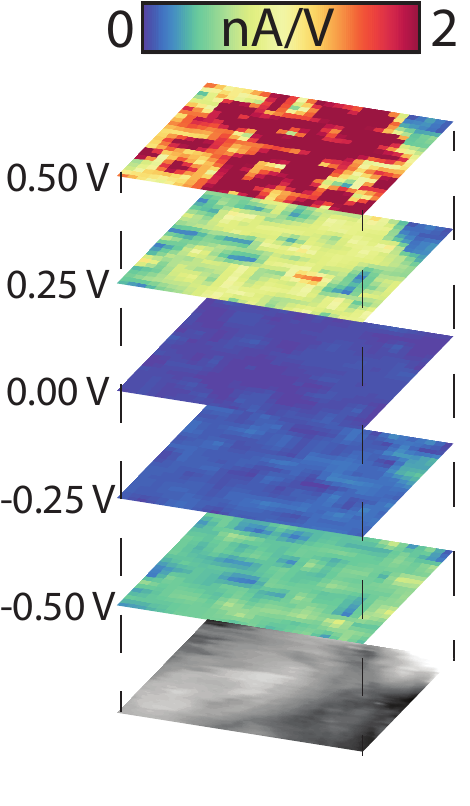}
\caption{d$I$/d$V$ maps at different bias voltage set points. The lateral position is aligned with the topography image at the bottom (80 $\times$ 80 nm). The height difference is clearly visible in the LDOS maps for $V~\geq~0.25$ V.}
\label{4ucBBO_LDOSMap}
\end{figure}

In Fig.~\ref{4ucBBO_LDOSMap} the spatially resolved local density of states (LDOS) for the 4-unit-cell-thick BBO film is shown for different bias voltages. Around the Fermi energy, no contrast is observed, similar as for the 10-unit-cell-thick BBO film (see Fig. 3(d) in the main text). Only for $V~>~0$~V, clear correlations are observed between the LDOS maps and the topography (bottom most image, same as inset of Fig.~4(a) in the main text). The higher region (see red dot in the inset of Fig.~4(a) in the main text) has a significantly lower LDOS than the lower region (blue dot).

\subsection{Band gap determination}
\begin{figure*}
\includegraphics[clip,keepaspectratio,width=0.8\textwidth]{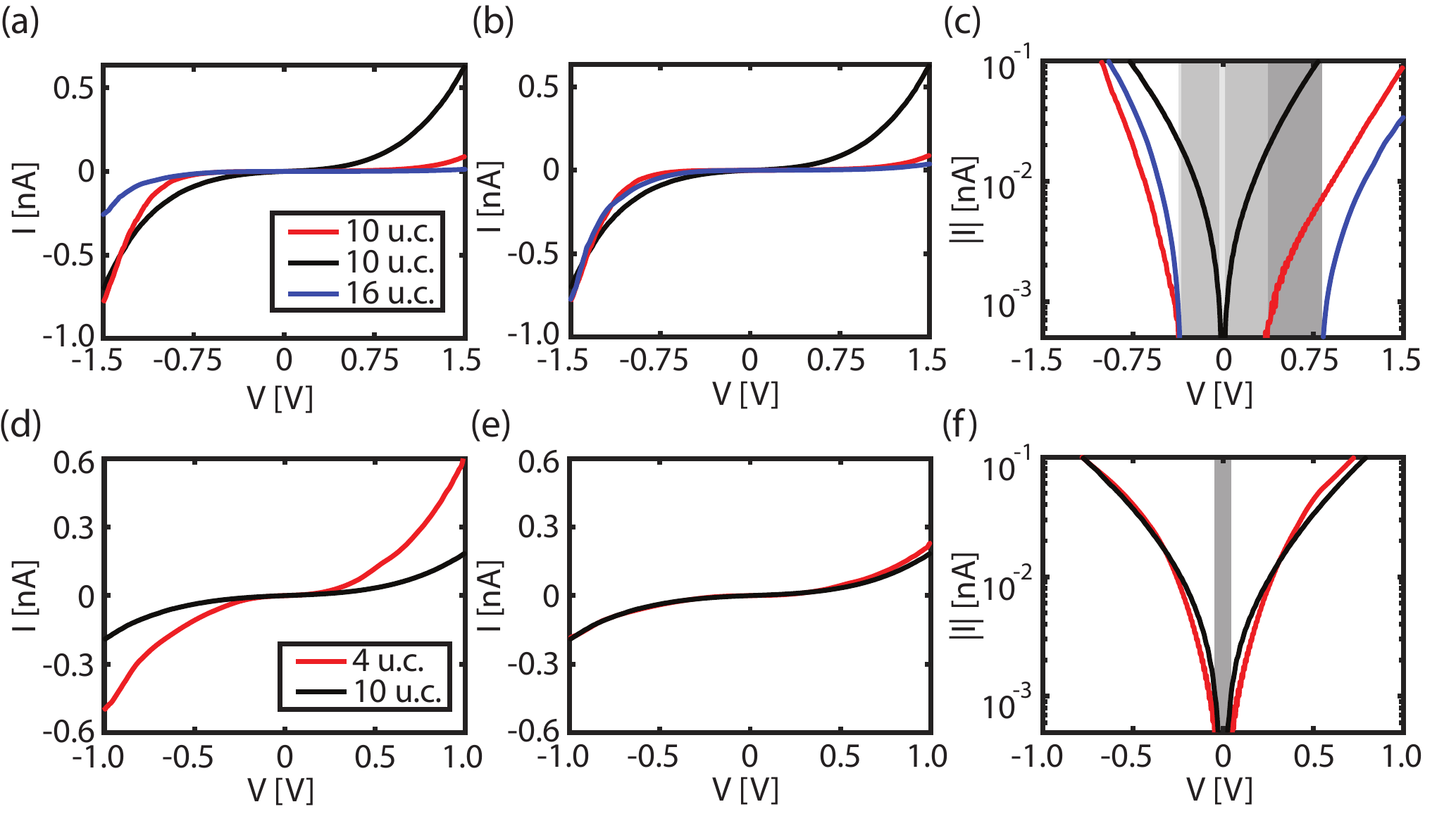}
\caption{(a-c) $I(V)$ curves recorded on the 10- and 16-unit-cell-thick BBO sample. (a) The original $I(V)$ curves. (b) The corrected $I(V)$ curves, scaled to the set point used for the \textit{I(V)} curse obtained on the 10 u.c. BBO film. (c) $I(V)$ curves on semi-logarithmic scale, the gray-colored bars represent the band gap of the various measurements. (d-f) $I(V)$ curves recorded on the 4 and 10~u.c. BBO films. (d) The original $I(V)$ curves. (e) The corrected $I(V)$ curves. (f) $I(V)$ curves on the semi-logarithmic scale.}
\label{BandgapFig}
\end{figure*}

In order to properly compare the band gap measured on the different samples, the $I(V)$ curves are first scaled to the same set point. In Fig.~\ref{BandgapFig}(a-c) the 10- and 16-unit-cell-thick BBO films are compared. Since it was not possible to use the same scan settings on both surfaces, the $I(V)$ spectrum of the 16~u.c. BBO film is scaled to the set point used for the $I(V)$ curves obtained on the 10~u.c. BBO film ($I$~=~600~pA and $V$~=~-1.5~V, Fig.~\ref{BandgapFig}(b)). The size of the band gap~($E_\textrm{G}$) is determined for all measurements by plotting the corrected $I(V)$ spectra on a semi-logarithmic scale (see Fig.~\ref{BandgapFig}) and, subsequently, taking the average voltage separation between the conduction band and valence band current onsets at the lowest detectable current (detection limit approximately 500~fA)~\cite{Feenstra1987,Ebert2011,Herbert2013}. 

From this measure a difference in the band gap is obtained between the 10 and 16~u.c. BBO films (see Fig.~\ref{BandgapFig}(c)). A similar approach is used to compare the 4 and 10~u.c. BBO samples, presented in Fig.~\ref{BandgapFig}(d-f). The obtained $I(V)$ curve on the 10-unit-cell-thick BBO film is scaled with respect to the $I(V)$ curve measure on the 4-unit-cell thick BBO film ($I$ = 400 pA and $V$ = 1 V). Although the absolute value for the band gap extracted from STS spectra depends slightly on the chosen scan parameters, the increasing trend between the thickness and the size of the band gap (Fig. 5 in the main text) remains unaffected.

\subsection{Topography and spectroscopy}
\begin{figure*}
\includegraphics[clip,keepaspectratio,width=0.8\textwidth]{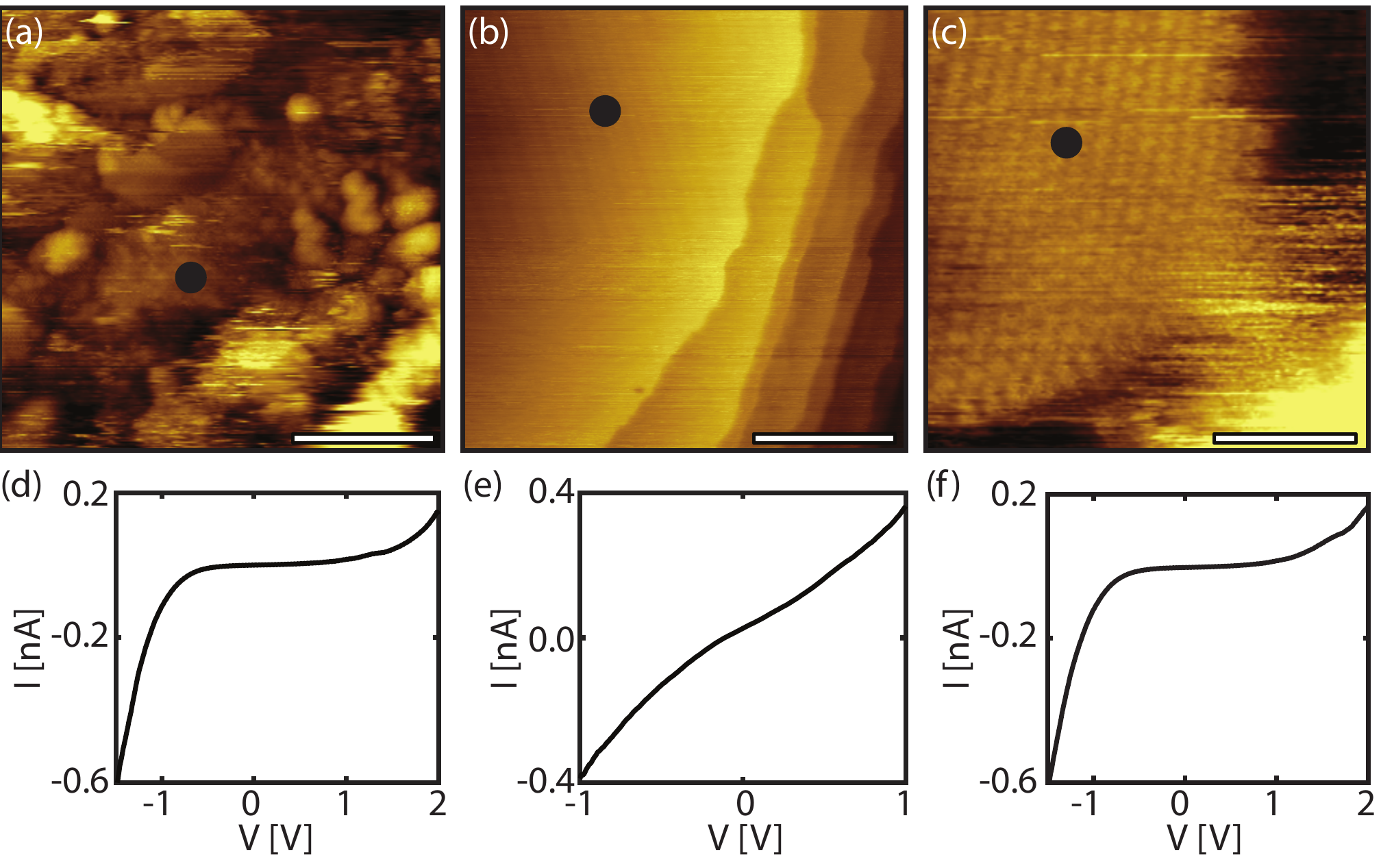}
\caption{Topography and spectroscopy data on 10-unit-cell-thick BBO films. (a) Topography of the BBO surface (200 $\times$ 200 nm, scale bar is 50 nm). (b) Topography image of Au(111) to show that the tip was in a good condition (100 $\times$ 100 nm, scale bar is 25 nm). (c) Topography scan of a 10 u.c. BBO film after scanning the Au(111) surface (20 $\times$ 20 nm, scale bar 5 nm). The BBO surface reconstruction is visible. (d-f) The corresponding \textit{I(V)} curves measured on the positions marked in panels (a-c), respectively.}
\label{SequenceFig}
\end{figure*}

Some of the topography images look a little bit scratchy, even though the measurements are repeatedly performed with different tips and scanners (the measurements are performed in the Nanoprobe STM which contains four indepently operating scanners). In order to exclude the possibility that the tips are contaminated, topography and spectroscopy data is presented in Fig.~\ref{SequenceFig}, taken before and after scanning on Au(111). 

In Fig.~\ref{SequenceFig}(a), a topography scan on a 10-unit-cell-thick BBO sample is presented. The topography looks a bit scratchy, but terraces are still visible. The measured $I(V)$ curve is displayed in Fig.~\ref{SequenceFig}(d) and shows the presence of a band gap. After the measurement on the BBO film, the sample was replaced by Au(111) (see Fig.~\ref{SequenceFig}(b)). Several steps are visible and on the terrace the herringbone reconstruction is present \cite{Woell1989}, indicating that the quality of the tip is good. Also, a metallic spectrum is measured in the $I(V)$ measurement, presented in Fig.~\ref{SequenceFig}(e). 

Subsequently, the 10-unit-cell-thick BBO film is scanned. A higher quality topography scan is obtained, see Fig.~\ref{SequenceFig}(c), with a similar spectroscopy measurement (Fig.~\ref{SequenceFig}(f)) as initially observed on BBO (Fig.~\ref{SequenceFig}(d)). Although the same tip and tunneling parameters are used, the image quality is slightly improved implicating that the scratchy topography appearance is not caused by the tip quality but reflects the state of the surface. Furthermore, the topography image of Fig.~\ref{SequenceFig}(a) (and also of Fig. 3 and 4 of the main text) look very similar to the topography images scanned with the AFM in Fig.~\ref{AFM_4uc} and Fig.~\ref{AFM_10uc}, suggesting that it is a not a tip artifact that is measured.

\newpage

\end{document}